\documentclass[prc,superscriptaddress,noshowpacs,unsortedaddress,twocolumn,showpacs,preprintnumbers,amsmath,amssymb]{revtex4-1}

\usepackage[dvipdfmx]{graphicx}
\usepackage{amsmath,amssymb,times}
\usepackage{color}
\usepackage{ulem}
\usepackage{bm}
\usepackage{here}

\newcommand{\beq}{\begin{equation}}
\newcommand{\eeq}{\end{equation}}
\newcommand{\bea}{\begin{eqnarray}}
\newcommand{\eea}{\end{eqnarray}}

\begin{document}

\title{Slope parameters determined from CREX and PREX2
}

\author{Shingo~Tagami}
\affiliation{Department of Physics, Kyushu University, Fukuoka 819-0395, Japan}

\author{Tomotsugu~Wakasa}
\affiliation{Department of Physics, Kyushu University, Fukuoka 819-0395, Japan}

\author{Masanobu~Yahiro}
\email[]{orion093g@gmail.com}
\affiliation{Department of Physics, Kyushu University, Fukuoka 819-0395, Japan}

\begin{abstract}
\begin{description}
\item[Background]
Very lately, the CREX group presents a skin value $\Delta R_{\rm skin}^{48}({\rm CREX}) 
=0.121 \pm 0.026\ {\rm (exp)} \pm 0.024\ {\rm (model)}=0.071\sim 0.171$~fm. 
Meanwhile, the PREX group reported a skin value 
$\Delta R_{\rm skin}^{208}({\rm PREX2}) = 0.283\pm 0.071=0.212 \sim 0.354$~fm. 
In our previous paper, we determined both the $L$--$\Delta R_{\rm skin}^{48}$ relation  and the $L$--$\Delta R_{\rm skin}^{208}$ one, 
using 206 EoSs, where $L$ is a slope parameter. 
\item[Purpose]
We  determine $L$ from $\Delta R_{\rm skin}^{48}({\rm CREX})$ and $\Delta R_{\rm skin}^{208}({\rm PREX2}) $, using 207 EoSs.
\item[Results]
The $\Delta R_{\rm skin}^{48}({\rm CREX})$ yields $L({\rm CREX})=0 \sim 51$~MeV and  the $\Delta R_{\rm skin}^{208}({\rm PREX2}) $ 
does $L({\rm PREX2})=76 \sim 165$~MeV. 
\item[Conclusion]
There is no overlap between  $L({\rm CREX})$ and $L({\rm PREX2})$. This is a big problem to be solved. 
\end{description}
 \end{abstract}

\maketitle

\section{Introduction and Conclusion}
\label{Sec:Introduction}

{\it Background on experiments:} 

Horowitz, Pollock and Souder proposed a direct measurement 
for neutron skin thickness $\Delta R_{\rm skin}=r_{\rm n}-r_{\rm p}$~\cite{PRC.63.025501}, 
where $r_{\rm p}$ and $r_{\rm n}$ are proton and neutron radii, respectively.  
This direct measurement consists of parity-violating and elastic electron scattering. 
In fact, 
the PREX collaboration has reported a new value, 
\begin{equation}
\Delta R_{\rm skin}^{208}({\rm PREX2}) = 0.283\pm 0.071= 0.212 \sim 0.354
\,{\rm fm}, 
\label{Eq:Experimental constraint 208}
\end{equation}
combining the original Lead Radius EXperiment (PREX)  result \cite{PRL.108.112502,PRC.85.032501} 
with the updated PREX2 result \cite{Adhikari:2021phr}. 
For $^{48}$Ca, the CREX group presents~\cite{CREX:2022kgg} 
\bea
\Delta R_{\rm skin}^{48}({\rm CREX}) 
&=&0.121 \pm 0.026\ {\rm (exp)} \pm 0.024\ {\rm (model)}
\notag \\
&=&0.071\sim 0.171~{\rm fm} .  
\eea
Note that the direct values are obtained from a single momentum transfer $q$. 
As an {\it ab initio} method for $^{48}$Ca,
we should consider the coupled-cluster  (CC) method~\cite{Hagen:2013nca,Hagen:2015yea} with chiral interaction. 
The CC result
$\Delta R_{\rm skin}^{48}({\rm CC})=0.12  \sim 0.15~ {\rm fm}$~\cite{Hagen:2015yea}  
is consistent with $\Delta R_{\rm skin}^{48}({\rm CREX})$.

As an indirect measurement on $\Delta R_{\rm skin}$, the high-resolution $E1$ polarizability 
experiment ($E1$pE)  was made 
for $^{208}$Pb~\cite{Tamii:2011pv} and $^{48}$Ca~\cite{Birkhan:2016qkr} in RCNP. 
The results are
\bea
\Delta R_{\rm skin}^{208}(E1{\rm pE}) &=&0.156^{+0.025}_{-0.021}=0.135 \sim 0.181~{\rm fm}, 
\label{Eq:skin-Pb208-E1}
\\
\Delta R_{\rm skin}^{48}(E1{\rm pE}) &=&0.14 \sim 0.20~{\rm fm}.  
\label{Eq:skin-Ca48-E1}
\eea
The $\Delta R_{\rm skin}^{48}({\rm CREX})$  is consistent with  $\Delta R_{\rm skin}^{48}(E1{\rm pE})$, 
 but $\Delta R_{\rm skin}^{208}({\rm PREX2})$ is not with $\Delta R_{\rm skin}^{208}(E1{\rm pE})$. 

Reaction cross section $\sigma_{\rm R}$ is a standard observable
to determine the matter radius $r_{\rm m}$ and the skin value $\Delta R_{\rm skin}$. 
High-accuracy data $\sigma_{\rm R}({\rm exp})$ are available for 
$^{42-51}$Ca + $^{12}$C scattering at 280~MeV per nucleon~\cite{Tanaka:2019pdo}. 
The chiral (Kyushu) $g$-matrix folding model~\cite{Toyokawa:2017pdd} 
yields $\Delta R_{\rm skin}^{48}({\rm exp})=0.105 \pm 0.06$~fm  
from the data~\cite{Takechi:2020snn}. 
The result is consistent with  $\Delta R_{\rm skin}^{48}({\rm CREX})$.
High-accuracy data $\sigma_{\rm R}({\rm exp})$ are available also for 
p + $^{208}$Pb scattering in $21  \le E_{\rm lab} \le 180$~MeV~\cite{PRC.12.1167,NPA.653.341,PRC.71.064606}. 
The chiral (Kyushu) $g$-matrix folding model yields 
$\Delta R_{\rm skin}^{208}({\rm exp})=0.278 \pm 0.035$~fm~\cite{Tagami:2020bee}.  
The value is  consistent with   $\Delta R_{\rm skin}^{208}({\rm PREX2})$. 

{\it Matter:}

Among the basic physical quantities that  determine the equation of state (EoS) of nuclear systems, the symmetry energy ($S_{\rm sym}$) and its dependence on the nucleon density ($\rho$) are receiving a lot of attention, 
because of their critical role in shaping the structure of nuclei and neutron stars (NSs)~\cite{Horowitz:2000xj,Lattimer:2014sga, Baldo:2016jhp, Oertel:2016bki,Piekarewicz:2019ahf,Zhang:2020azr,Burgio:2021vgk}. 
Many predictions on the symmetry energy$S_{\rm sym}(\rho)$ have been made so far  
by taking several experimental and observational constraints on $S_{\rm sym}(\rho)$ and their combinations. 
The $\Delta R_{\rm skin}^{208}(E1{\rm pE})$ and the $\Delta R_{\rm skin}^{48}({\rm CREX})$ are  
the most important experimental constraint on the slope parameter $L$, since a strong correlation between 
$r_{\rm skin}^{208}$ and $L$ is well known~\cite{RocaMaza:2011pm,Brown:2013mga}. 
In this paper, we will show a strong correlation between  $r_{\rm skin}^{48}$ and $L$ and between 
$r_{\rm skin}^{208}$ and $L$ from 207 EoSs. 

As an essential constraint on the EoS from astrophysics, one may take 
$M = 1.97 \pm 0.04{\rm M}_{\rm sun}$~\cite{Demorest:2010bx}, where $M$ is the mass of NS.  
For a pulsar in a binary system, detection of the general relativistic Shapiro delay allows us to determine $M$. 

Usually, the $S_{\rm sym}$ is expanded into 
\bea
S_{\rm sym}(\rho)=J + \frac{L (\rho-\rho_0)}{3\rho_0}+\frac{K_{\rm sym} (\rho-\rho_0)^2}{18\rho_0^2}+ 
\cdots .
\eea
in terms of the nuclear density $\rho$ around  the saturation density  $\rho_0$. 
For the $S_{\rm sym}(\rho)$, at the present stage, a major aim is to determine  $L$ at $\rho=\rho_0$. 
The symmetry energy $S_{\rm sym}(\rho)$ cannot be measured by experiment directly. 
In place of  $S_{\rm sym}(\rho)$, the neutron-skin thickness $r_{\rm skin}$ is measured to determine 
$L$. This subject  is currently under  experimental investigation for $^{208}$Pb and $^{48}$Ca nuclei at Jefferson Lab~\cite{Abrahamyan:2012gp,Tagami:2020shn,PREX:2021umo}. 

\begin{figure}[h]
\begin{center}
 \includegraphics[width=0.40\textwidth,clip]{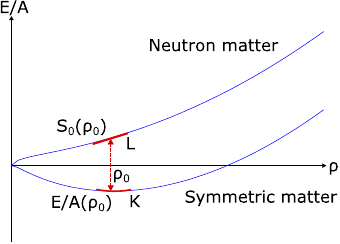}
  \caption{Illustration on the physics meaning of the slope parameter $L$ and the incompressibility $K$.
    }
 \label{Fig-E-rho}
\end{center}
\end{figure}

Figure \ref{Fig-E-rho} shows the binding energy per nucleon $E/A$ as a function of $\rho$ and 
$\delta=(N-Z)/A$. 
The  $E/A$ is expanded into
\bea
\frac{E(\rho,\delta)}{A}=S_0(\rho) +S_{\rm sym}(\rho)\delta^2+\cdots
\eea 
in terms of $\delta$; note that $J=S_0(\rho_0)$. This equation shows the relation between $E/A(\rho,\delta)$ 
and $S_{\rm sym}(\rho)$. Two curves correspond to the symmetric-nuclear matter with $\delta=0$ and 
pure-neutron matter with  $\delta=1$, respectively. 
This figure shows that $L$ is a slope for pure-neutron matter at  $\rho=\rho_0$, 
because $dS_0(\rho)/d\rho$  is zero at  $\rho=\rho_0$.

We  accumulate the 205 EoSs from Refs.~\cite{Akmal:1998cf,RocaMaza:2011pm,Ishizuka:2014jsa,Gonzalez-Boquera:2017rzy,D1P-1999,Gonzalez-Boquera:2017uep,Oertel:2016bki,Piekarewicz:2007dx,Lim:2013tqa,Sellahewa:2014nia,Inakura:2015cla,Fattoyev:2013yaa,Steiner:2004fi,Centelles:2010qh,Dutra:2012mb,Brown:2013pwa,Brown:2000pd,Reinhard:2016sce,Tsang:2019ymt,Ducoin:2010as,Fortin:2016hny,Chen:2010qx,Zhao:2016ujh,Zhang:2017hvh,Wang:2014rva,Lourenco:2020qft,Matsuzaki:2022spe,Matsuzaki:2021hdm} 
 in which $\Delta R_{\rm skin}^{208}$ and/or $L$ is presented, since 
a strong correlation between $\Delta R_{\rm skin}^{208}$ and a slope parameter $L$ is shown. 
In the 205 EoSs of Table I, the number of Gogny EoSs is much smaller than that of Skyrme EoSs.
In Ref.~\cite {TAGAMI2022105155}, we constructed D1MK and D1PK. 
Eventually, we get the 207 EoSs, as shown in Table I.  
The correlation is more reliable when the number of EoSs is larger. 
For this reason,  we  take  the 207 EoSs.

Among the 207 EoSs, APR of Ref.~\cite{Akmal:1998cf}  is the most reliable EoS, since 
they calculated properties of dense symettric -nuclear and pure-nuclear matter and 
the structure of neutron stars  with 
the variational chain summation methods for the Argonne v18
two-nucleon interaction plus the Urbana three nucleon interaction. 
The methods are hard to calculate. In fact, Steiner {\it el.al.} have fit bulk properties of APR 
for symmetric-nuclear and 
neutron matter  with a Skyrme-like Hamiltonian~\cite{Steiner:2004fi}. The EoS is called NRAPR. 
Brown and Schwenk modified NRAPR slightly~\cite{Brown:2013pwa}. 
The EoS is called  NRAPR-B in this paper.
Tsang  {\it el.al.} made further modification for NRAPR-B~\cite{Tsang:2019ymt}. 
The EoS is called  NRAPR-T in this paper.

\begin{figure}[H]
\begin{center}
 \includegraphics[width=0.45\textwidth,clip]{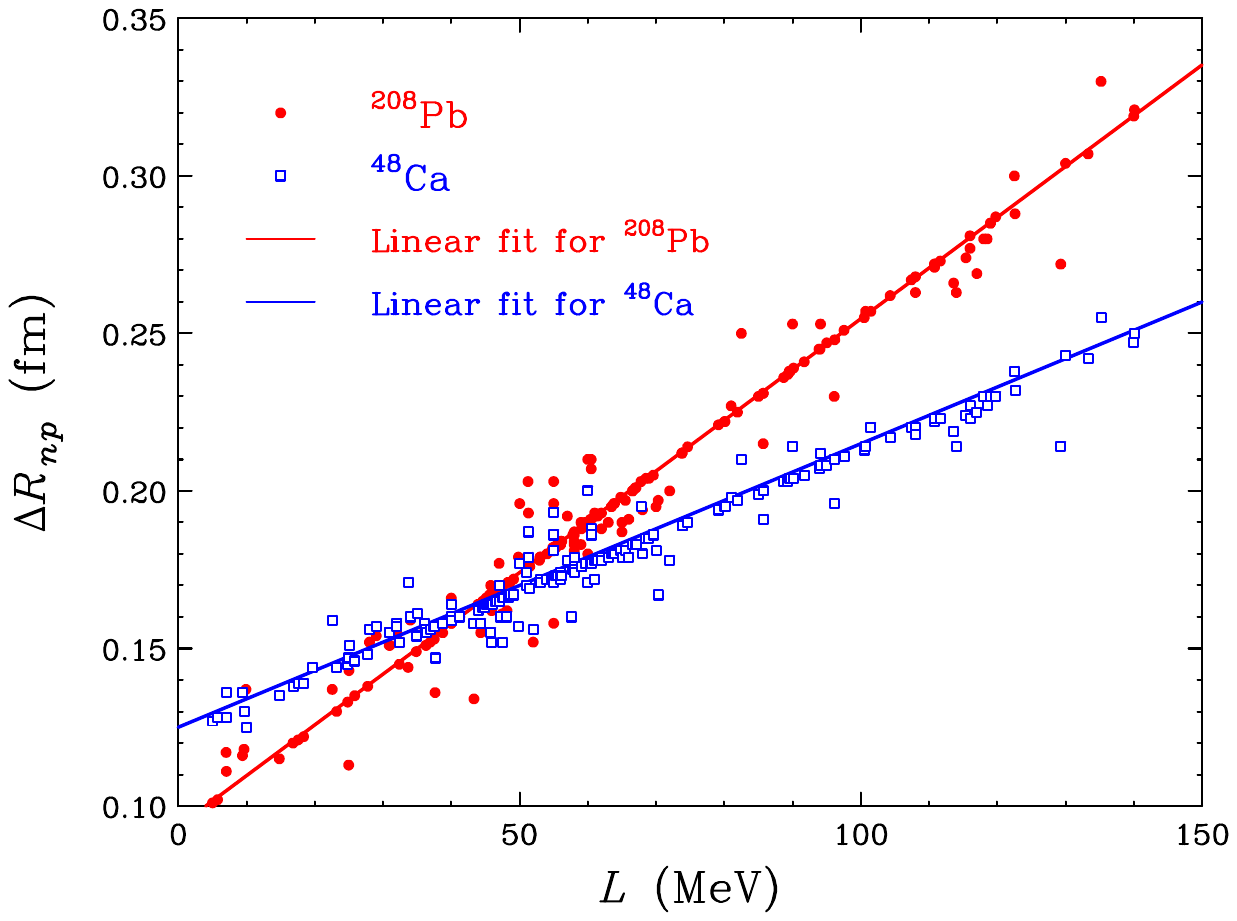}
  \caption{ 
Skin value $\Delta R$ as a function of $L$ for $^{208}$Pb and $^{48}$Ca.    
Two straight line show  Eq.~\eqref{final-48-skin-L} 
and Eq.~\eqref{Eq:skin-L}, respectively. Dots denote  207 EoSs for $^{208}$Pb and $^{48}$Ca. 
    }
 \label{Fig-skin-L-1}
\end{center}
\end{figure}

The $\Delta R_{\rm skin}^{208}({\rm PREX2})$ and $\Delta R_{\rm skin}^{48}({\rm CREX})$ is most reliable, 
 and provides  $L$ of nuclear matter. 
 In fact, using 207 EoSs of Table I, we found the $L$-$\Delta R_{\rm skin}^{48}$  relation~\cite{TAGAMI2022105155} as 
\bea
\Delta R_{\rm skin}^{48}=0.0009 L + 0.125 > 0.125~{\rm fm}
\label{final-48-skin-L}
\eea
with a high correlation coefficient $R=0.98$, because of $L >0$. Equation \eqref{final-48-skin-L} indicates that 
the lower limit of $\Delta R_{\rm skin}^{48}$ is 0.125~fm. 
The same derivation is possible the $L$-$\Delta R_{\rm skin}^{208}$ relation: 
\bea
L=620.39~\Delta R_{\rm skin}^{208}-57.963 
\label{Eq:skin-L}
\eea
has $R=0.99$. 
The two relations are visualized by  Fig.~\ref{Fig-skin-L-1}. 

\begin{figure}[h]
\begin{center}
 \includegraphics[width=0.45\textwidth,clip]{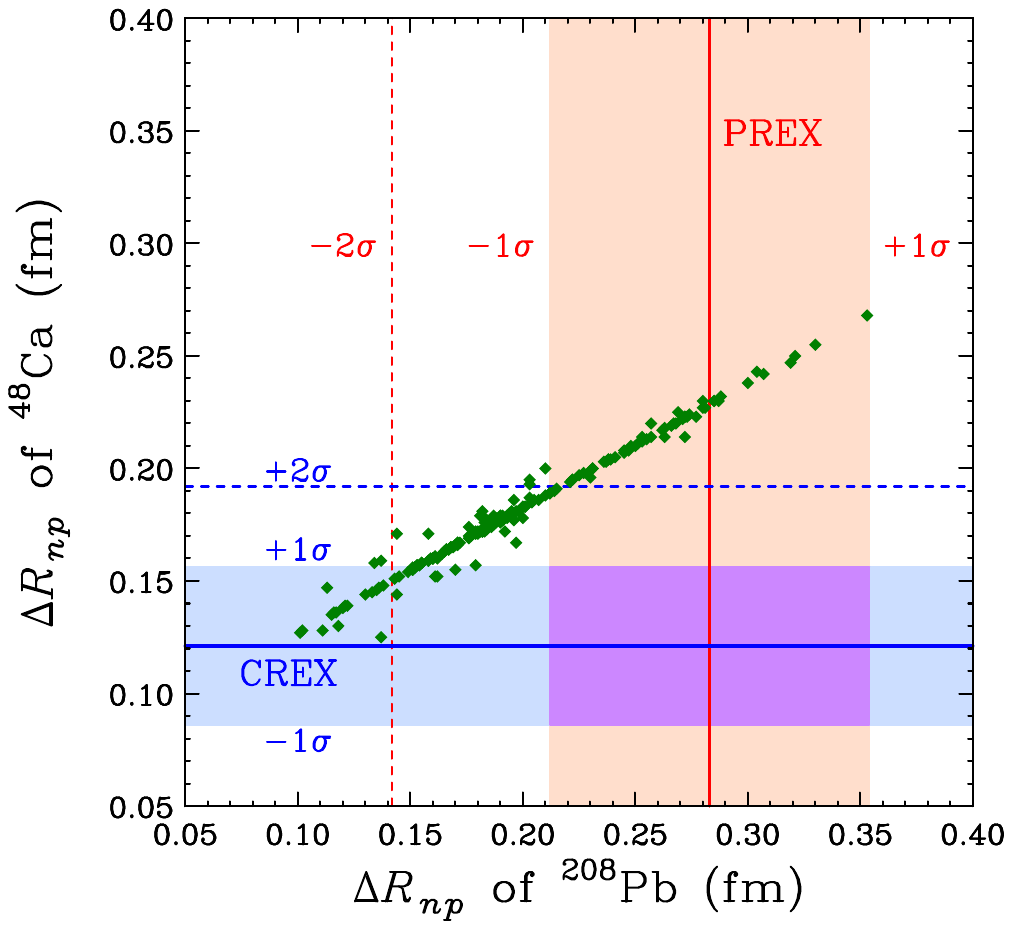}
  \caption{ Relation between $\Delta R_{\rm skin}^{208}$ and $\Delta R_{\rm skin}^{48}$.
    }
 \label{Fig-208-48}
\end{center}
\end{figure}

{\it Results:}
Figure \ref{Fig-208-48} shows the relation between $\Delta R_{\rm skin}^{208}$ and $\Delta R_{\rm skin}^{48}$.
The 207 EoSs do not satisfy both $\Delta R_{\rm skin}^{208}({\rm PREX2}) $ and $\Delta R_{\rm skin}^{48}({\rm CREX})$ in the one-$\sigma$ level. 
If one considers the two-$\sigma$ level of  $\Delta R_{\rm skin}^{208}({\rm PREX2}) $ and $\Delta R_{\rm skin}^{48}({\rm CREX})$, one may find that some EoSs satisfy $\Delta R_{\rm skin}^{208}({\rm PREX2}) $ and $\Delta R_{\rm skin}^{48}({\rm CREX})$
  
The $L$--$\Delta R_{\rm skin}^{48}$ realtion of Eq.~\eqref{final-48-skin-L} yields 
$L({\rm CREX})=0 \sim 51$~MeV and  
the $L$--$\Delta R_{\rm skin}^{208} of $Eq.~\eqref{Eq:skin-L} does $L({\rm PREX2})=76 \sim 165$~MeV. 
There is no overlap between them. 
Note that the value of $L$ should be positive.  
The later value is consistent with  $L = 76 \sim 172$ MeV of Ref.~\cite{RocaMaza:2011pm}.
Reed {\it et al.} \cite{Reed:2021nqk} 
report a value of the slope parameter $L= 106 \pm 37=69	\sim 143 ~{\rm MeV}$.

{\it Conclusion:}There is no overlap between  $L({\rm CREX})$ and $L({\rm PREX2})$  in the one-$\sigma$ level, 
as shown in Fig.~\ref{Fig-208-48}. This is a big problem to be solved. 
As another problem, we can say that APR \cite{Akmal:1998cf,Brown:2000pd, Ishizuka:2014jsa} as the most reliable EoS is ruled out.

\section*{Acknowledgements}
We thank Prof. Matsuzaki   for his comments.

\bibliography{Folding-v13}

\onecolumngrid

\squeezetable
\begin{table}[h]
\caption{Properties of 207 EoSs (1).    
For the 207 EoSs, both  $\Delta R_{\rm skin}^{208}$and $L$ are obtained self-consistently; 
the starting $\Delta R_{\rm skin}^{208}$-$L$ relation is  determined from the EoSs in which 
both $\Delta R_{\rm skin}^{208}$ and $L$ are presented. For the 207 EoSs, we determine the relation 
between $\Delta R_{\rm skin}^{208}$ and $\Delta R_{\rm skin}^{48}$ 
self-consistently, where  
the starting $\Delta R_{\rm skin}^{208}$-$\Delta R_{\rm skin}^{48}$ relation is  determined from the EoSs in which 
both $\Delta R_{\rm skin}^{208}$ and $\Delta R_{\rm skin}^{48}$ are presented. 
The symbol $^\dagger$ shows results of  self-consistent calculations.}
\label{table-I}
\begin{tabular}{c|ccccccc|cc}
\hline
 & m*/m & K &J & L & Ksym & Rskin-208 & Rskin-48 & Refs. \cr
\hline	      
APR, E0019& & 266.000  & 32.600  & 57.600  &  & 0.160   & 0.160$^\dagger$  
& \cite{Akmal:1998cf,Brown:2000pd, Ishizuka:2014jsa}   & \cr
BHF-1 &  & 195.500  & 34.300  & 66.550  & -31.300  & 0.200$^\dagger$  & 0.183$^\dagger$  & 
\cite{Ducoin:2010as}  & \cr
BSk14 & 0.800  & 239.380  & 30.000  & 43.910  & -152.030  & 0.164$^\dagger$  & 0.162$^\dagger$  & 
\cite{Ducoin:2010as,Dutra:2012mb}  &\cr 
BSk16 & 0.800  & 241.730  & 30.000  & 34.870  & -187.390  & 0.149$^\dagger$  & 0.154$^\dagger$  & 
\cite{Ducoin:2010as,Dutra:2012mb} & \cr
BSk17 & 0.800  & 241.740  & 30.000  & 36.280  & -181.860  & 0.151$^\dagger$  & 0.156$^\dagger$  & 
\cite{Ducoin:2010as,Dutra:2012mb}  & \cr
BSk20 & 0.800  & 241.400  & 30.000  & 37.400  & -136.500  & 0.153$^\dagger$  & 0.157$^\dagger$  &  \cite{Fortin:2016hny,Dutra:2012mb}  & \cr
BSk21 & 0.800  & 245.800  & 30.000  & 46.600  & -37.200  & 0.168$^\dagger$  & 0.165$^\dagger$  &  \cite{Fortin:2016hny,Dutra:2012mb}  & \cr
BSk22 &  & 245.900  & 32.000  & 68.500  & 13.000  & 0.204$^\dagger$  & 0.185$^\dagger$  &  
\cite{Fortin:2016hny}  & \cr
BSk23 &  & 245.700  & 31.000  & 57.800  & -11.300  & 0.186$^\dagger$  & 0.175$^\dagger$  & 
\cite{Fortin:2016hny,Ishizuka:2014jsa}   & \cr
BSk24 &  & 245.500  & 30.000  & 46.400  & -37.600  & 0.168$^\dagger$  & 0.165$^\dagger$  & 
\cite{Fortin:2016hny}   & \cr
BSk25 &  & 236.000  & 29.000  & 36.900  & -28.500  & 0.152$^\dagger$  & 0.156$^\dagger$   & 
\cite{Fortin:2016hny}   & \cr
BSk26 &  & 240.800  & 30.000  & 37.500  & -135.600  & 0.153$^\dagger$  & 0.157$^\dagger$  & 
\cite{Fortin:2016hny}   & \cr
BSR2 &  & 239.900  & 31.500  & 62.000  & -3.100  & 0.193$^\dagger$  & 0.179$^\dagger$  & 
\cite{Fortin:2016hny}   & \cr
BSR6 &  & 235.800  & 35.600  & 85.700  & -49.600  & 0.231$^\dagger$  & 0.200$^\dagger$  & 
\cite{Fortin:2016hny}   & \cr
D1 &  & 229.400  & 30.700  & 18.360  & -274.600  & 0.122$^\dagger$  & 0.139$^\dagger$  & 
\cite{Gonzalez-Boquera:2017uep}   & \cr
D1AS &  & 229.400  & 31.300  & 66.550  & -89.100  & 0.200$^\dagger$  & 0.183$^\dagger$  & 
\cite{Gonzalez-Boquera:2017uep}  & \cr
D1M & 0.746$^\ddagger$  & 224.958$^\ddagger$  & 28.552$^\ddagger$  & 24.966$^\ddagger$  & -133.692$^\ddagger$  & 0.113$^\ddagger$  & 0.147$^\ddagger$  & 
\cite{Gonzalez-Boquera:2017rzy}  & \cr
D1M* & 0.746$^\ddagger$  & 225.365$^\ddagger$  & 30.249$^\ddagger$  & 43.311$^\ddagger$  & -47.793$^\ddagger$  & 0.134$^\ddagger$  & 0.158$^\ddagger$  & 
\cite{Gonzalez-Boquera:2017rzy}   & \cr
D1MK & 0.746$^\ddagger$  & 225.400$^\ddagger$  & 33.000$^\ddagger$  & 55.000$^\ddagger$  & -37.275$^\ddagger$  & 0.158$^\ddagger$  & 0.171$^\ddagger$  & \cite{TAGAMI2022105155}  & \cr
D1N & 0.748$^\ddagger$  & 225.525$^\ddagger$  & 29.594$^\ddagger$  & 33.665$^\ddagger$  & -168.750$^\ddagger$  & 0.144$^\ddagger$  & 0.171$^\ddagger$  & 
\cite{Gonzalez-Boquera:2017rzy}  & \cr
D1P & 0.672$^\ddagger$  & 250.860$^\ddagger$  & 32.418$^\ddagger$  & 49.827$^\ddagger$  & -157.419$^\ddagger$  & 0.179$^\ddagger$  & 0.157$^\ddagger$  & 
\cite{Gonzalez-Boquera:2017uep}  & \cr
D1PK & 0.700$^\ddagger$  & 260.000$^\ddagger$  & 33.000$^\ddagger$  & 55.000$^\ddagger$  & -150.000$^\ddagger$  & 0.182$^\ddagger$  & 0.181$^\ddagger$  & \cite{TAGAMI2022105155}  & \cr
D1S & 0.697$^\ddagger$  & 202.856$^\ddagger$  & 31.125$^\ddagger$  & 22.558$^\ddagger$  & -241.797$^\ddagger$  & 0.137$^\ddagger$  & 0.159$^\ddagger$  & 
\cite{Gonzalez-Boquera:2017rzy,Inakura:2015cla}  & \cr
D2 & 0.738  & 209.300  & 31.130  & 44.850  &  & 0.165$^\dagger$  & 0.163$^\dagger$  & 
\cite{Gonzalez-Boquera:2017rzy}  & \cr
D250 &  & 249.900  & 31.570  & 24.820  & -289.400  & 0.133$^\dagger$  & 0.145$^\dagger$  & 
\cite{Gonzalez-Boquera:2017uep}   & \cr
D260 &  & 259.500  & 30.110  & 17.570  & -298.700  & 0.121$^\dagger$  & 0.139$^\dagger$  & 
\cite{Gonzalez-Boquera:2017uep}   & \cr
D280 &  & 285.200  & 33.140  & 46.530  & -211.900  & 0.168$^\dagger$  & 0.165$^\dagger$  & 
\cite{Gonzalez-Boquera:2017uep}   & \cr
D300 &  & 299.100  & 31.220  & 25.840  & -315.100  & 0.135$^\dagger$  & 0.146$^\dagger$  & 
\cite{Gonzalez-Boquera:2017uep}   & \cr
DD &  & 241.000  & 31.700  & 56.000  & -95.000  & 0.183$^\dagger$  & 0.173$^\dagger$  &  
\cite{Chen:2010qx}  & \cr
DD-F &  & 223.000  & 31.600  & 56.000  & -140.000  & 0.183$^\dagger$  & 0.173$^\dagger$  & 
\cite{Chen:2010qx}  & \cr
DD-ME1 &  & 245.000  & 33.100  & 55.000  & -101.000  & 0.203$^\dagger$  & 0.193$^\dagger$  & 
\cite{Chen:2010qx,Zhao:2016ujh,Ducoin:2010as,Ishizuka:2014jsa}  & \cr
DD-ME2 &  & 251.000  & 32.300  & 51.240  & -87.000  & 0.203$^\dagger$  & 0.187$^\dagger$  & 
\cite{Chen:2010qx,Zhao:2016ujh,Ducoin:2010as,Fortin:2016hny}  & \cr
DD-PC1 &  &  &  & 67.799  &  & 0.203$^\dagger$  & 0.195$^\dagger$  & 
\cite{Zhao:2016ujh,Ducoin:2010as}  & \cr
Ducoin &  & 240.200  & 32.760  & 55.300  & -124.700  & 0.182$^\dagger$  & 0.173$^\dagger$  & 
\cite{Ducoin:2010as}   & \cr
E0008(TMA) &  & 318.000  & 30.660  & 90.140  &  & 0.239$^\dagger$  & 0.204$^\dagger$  & 
\cite{Ishizuka:2014jsa}  & \cr
E0009 &  & 280.000  & 32.500  & 88.700  &  & 0.236$^\dagger$  & 0.203$^\dagger$  & 
\cite{Ishizuka:2014jsa,Oertel:2016bki}  & \cr
E0015 &  & 216.700  & 30.030  & 45.780  &  & 0.167  & 0.164  & 
\cite{Ishizuka:2014jsa}  & \cr
E0024 &  & 244.500  & 33.100  & 55.000  &  & 0.182$^\dagger$  & 0.172  & 
\cite{Ishizuka:2014jsa}  & \cr
E0025 &  & 211.000  & 31.600  & 107.400  &  & 0.267$^\dagger$  & 0.220$^\dagger$  & 
\cite{Ishizuka:2014jsa}  & \cr
E0036 &  & 281.000  & 36.900  & 110.800  &  & 0.272$^\dagger$  & 0.223$^\dagger$  & 
\cite{Ishizuka:2014jsa}  & \cr
es25 &  & 211.730  & 25.000  & 27.749$^\dagger$  &  & 0.138  & 0.148$^\dagger$  & 
\cite{Steiner:2004fi}   & \cr
es275 &  & 205.330  & 27.500  & 48.549$^\dagger$  &  & 0.171  & 0.167$^\dagger$  & 
\cite{Steiner:2004fi,Fortin:2016hny}  & \cr
es30 &  & 215.360  & 30.000  & 69.603$^\dagger$  &  & 0.205  & 0.186$^\dagger$  & 
\cite{Steiner:2004fi}  & \cr
es325 &  & 212.450  & 32.500  & 81.925$^\dagger$  &  & 0.225  & 0.197$^\dagger$  & 
\cite{Steiner:2004fi}  & \cr
es35 &  & 209.970  & 34.937  & 96.182$^\dagger$  &  & 0.248  & 0.210$^\dagger$  & 
\cite{Steiner:2004fi}  & \cr
FKVW &  & 379.000  & 33.100  & 80.000  & 11.000  & 0.222$^\dagger$  & 0.195$^\dagger$  & 
\cite{Chen:2010qx}   & \cr
FSU &  & 230.000  & 32.590  & 60.500  & -51.300  & 0.210  & 0.188$^\dagger$  & 
\cite{Fattoyev:2013yaa,Ducoin:2010as}  & \cr
FSUgold &  & 229.000  & 32.500  & 60.000  & -52.000  & 0.210  & 0.200  & 
\cite{Chen:2010qx,Oertel:2016bki,Piekarewicz:2007dx}    & \cr
FSUgold2.1 &  & 230.000  & 32.590  & 60.500  &  & 0.191$^\dagger$  & 0.177$^\dagger$  & 
\cite{Ishizuka:2014jsa,Oertel:2016bki}   & \cr
GM1 &  & 299.700  & 32.480  & 93.870  & 17.890  & 0.245$^\dagger$  & 0.207$^\dagger$  & 
\cite{Ducoin:2010as,Fortin:2016hny}   & \cr
GM3 &  & 239.900  & 32.480  & 89.660  & -6.470  & 0.238$^\dagger$  & 0.204$^\dagger$  & 
\cite{Ducoin:2010as}  & \cr
Gs &  & 237.570  & 31.384  & 89.304$^\dagger$  &  & 0.237  & 0.203$^\dagger$  & 
\cite{Steiner:2004fi}  & \cr
GSkI &  & 230.210  & 32.030  & 63.450  & -95.290  & 0.195$^\dagger$  & 0.180$^\dagger$  & 
\cite{Dutra:2012mb}   & \cr
GSkII & 0.790  & 233.400  & 30.490  & 48.630  & -157.830  & 0.171$^\dagger$  & 0.167$^\dagger$  & \cite{Dutra:2012mb}   & \cr
GT2 &  & 228.100  & 33.940  & 5.020  & -445.900  & 0.101$^\dagger$  & 0.127$^\dagger$  & 
\cite{Gonzalez-Boquera:2017uep}  & \cr
G$\sigma$ &  & 237.290  & 31.370  & 94.020  & 13.990  & 0.245$^\dagger$  & 0.208$^\dagger$  & 
\cite{Ducoin:2010as}  & \cr
HA &  & 233.000  & 30.700  & 55.000  & -135.000  & 0.182$^\dagger$  & 0.172$^\dagger$  & 
\cite{Chen:2010qx}  & \cr
HFB-17 &  &  &  & 36.300  &  & 0.151  & 0.155$^\dagger$  & 
\cite{RocaMaza:2011pm}   & \cr
HFB-8 &  &  &  & 14.800  &  & 0.115  & 0.135$^\dagger$  & 
\cite{RocaMaza:2011pm}  & \cr
HS(DD2) &  & 243.000  & 31.700  & 55.000  & -93.200  & 0.182$^\dagger$  & 0.172$^\dagger$  & 
\cite{Oertel:2016bki,Fortin:2016hny}  & \cr
IU-FSU &  & 231.200  & 31.300  & 47.200  & 28.700  & 0.160  & 0.160$^\dagger$  & 
\cite{Fattoyev:2013yaa, Oertel:2016bki}    & \cr
KDE0v1 & 0.740  & 227.540  & 34.580  & 54.690  & -127.120  & 0.181$^\dagger$  & 0.172$^\dagger$  & 
\cite{Dutra:2012mb}   & \cr
KDE0v1-B & 0.790  & 216.000  & 34.900  & 61.000  &  & 0.192  & 0.172  & 
\cite{Brown:2013pwa}  & \cr
\hline
\end{tabular}
\end{table} 
\setcounter{table}{0}
\begin{table}[h]
\caption{Properties of 207 EoSs (2).
}
\begin{tabular}{c|ccccccc|cc}
\hline
 & m*/m & K &J & L & Ksym & Rskin-208 & Rskin-48 & Refs. \cr
\hline
KDE0v1-T & 0.810  & 217.000  & 34.600  & 72.000  & -40.000  & 0.200  & 0.178  & 
\cite{Tsang:2019ymt}   & \cr
LNS & 0.830  & 210.780  & 33.430  & 61.450  & -127.360  & 0.192$^\dagger$  & 0.178$^\dagger$  & 
\cite{Dutra:2012mb,Ducoin:2010as}     & \cr
LS180 &  & 180.000  & 28.600  & 73.800  &  & 0.212$^\dagger$  & 0.189$^\dagger$  & 
\cite{Dutra:2012mb,Ishizuka:2014jsa,Oertel:2016bki}   & \cr
LS220 &  & 220.000  & 28.600  & 73.800  &  & 0.212$^\dagger$  & 0.189$^\dagger$  & 
\cite{Dutra:2012mb,Ishizuka:2014jsa,Oertel:2016bki}   & \cr
LS375 &  & 375.000  & 28.600  & 73.800  &  & 0.212$^\dagger$  & 0.189$^\dagger$  & 
\cite{Dutra:2012mb,Ishizuka:2014jsa,Oertel:2016bki}  & \cr
Ly5 &  & 229.940  & 32.010  & 45.243$^\dagger$  &  & 0.166  & 0.164$^\dagger$  & 
\cite{Steiner:2004fi}   & \cr
M3Y-P6 &  & 239.700  & 32.100  & 44.600  & -165.300  & 0.165$^\dagger$  & 0.163$^\dagger$  & 
\cite{Inakura:2015cla,Lim:2013tqa}  & \cr
M3Y-P7 &  & 254.700  & 31.700  & 51.500  & -127.800  & 0.176$^\dagger$  & 0.169$^\dagger$  & 
\cite{Inakura:2015cla,Fortin:2016hny}  & \cr
MSk3 & 1.000  & 233.250  & 28.000  & 7.040  & -283.520  & 0.111  & 0.128  &  
\cite{Dutra:2012mb,Zhao:2016ujh}   & \cr
MSk6 & 1.050  & 231.170  & 28.000  & 9.630  & -274.330  & 0.118  & 0.130  &  
\cite{Dutra:2012mb,Zhao:2016ujh}   & \cr
MSk7 & 1.050  & 385.360  & 27.950  & 9.400  & -274.630  & 0.116  & 0.136$^\dagger$  & 
\cite{RocaMaza:2011pm,Dutra:2012mb}   & \cr
MSL0 & 0.800  & 230.000  & 30.000  & 60.000  & -99.330  & 0.180  & 0.171$^\dagger$  & 
\cite{Chen:2010qx,Dutra:2012mb,Wang:2014rva}  & \cr
NL1 &  & 212.000  & 43.500  & 140.000  & 143.000  & 0.319  & 0.247  & 
\cite{Chen:2010qx,Zhao:2016ujh}   & \cr
NL2 &  & 401.000  & 44.000  & 130.000  & 20.000  & 0.304  & 0.243  & 
\cite{Chen:2010qx,Zhao:2016ujh}   & \cr
NL3 &  & 271.000  & 37.300  & 118.000  & 100.000  & 0.280  & 0.230  & 
\cite{Chen:2010qx,Fattoyev:2013yaa,Centelles:2010qh,Piekarewicz:2007dx,Ducoin:2010as,Fortin:2016hny} 
 & \cr
NL3* &  &  &  & 119.769$^\dagger$  &  & 0.287  & 0.230  & 
\cite{Zhao:2016ujh}  & \cr
NL3$\omega \rho$ &  & 271.600  & 31.700  & 55.500  & -7.600  & 0.183$^\dagger$  & 0.173$^\dagger$  & \cite{Fortin:2016hny}  & \cr
NL4 &  & 270.350  & 36.239  & 111.649$^\dagger$  &  & 0.273  & 0.223$^\dagger$  & 
\cite{Steiner:2004fi}   & \cr
NL-SH &  & 356.000  & 36.100  & 114.000  & 80.000  & 0.263  & 0.214  & 
\cite{Chen:2010qx,Zhao:2016ujh}   & \cr
NL$\rho$ &  & 240.000  & 30.300  & 85.000  & 3.000  & 0.230$^\dagger$  & 0.199$^\dagger$  & 
\cite{Chen:2010qx}  & \cr
NL$\omega \rho$(025) &  & 270.700  & 32.350  & 61.050  & -34.360  & 0.192$^\dagger$  & 0.178$^\dagger$  & \cite{Ducoin:2010as}  & \cr
NRAPR & 0.690  & 225.700  & 32.787  & 59.630  & -123.320  & 0.190  & 0.177$^\dagger$  & 
\cite{Steiner:2004fi,Dutra:2012mb}   & \cr
NRAPR-B & 0.850  & 225.000  & 35.100  & 61.000  &  & 0.193  & 0.178  & 
\cite{Brown:2013pwa}  & \cr
NRAPR-T & 0.730  & 221.000  & 34.100  & 70.000  & -46.000  & 0.195  & 0.181  & 
\cite{Tsang:2019ymt}  & \cr
PC-F1 &  & 255.000  & 37.800  & 117.000  & 75.000  & 0.269  & 0.225  & 
\cite{Chen:2010qx,Zhao:2016ujh}  & \cr
PC-F2 &  & 256.000  & 37.600  & 116.000  & 65.000  & 0.281$^\dagger$  & 0.227$^\dagger$  & 
\cite{Chen:2010qx,Zhao:2016ujh,Wang:2014rva}  & \cr
PC-F3 &  & 256.000  & 38.300  & 119.000  & 74.000  & 0.285$^\dagger$  & 0.230$^\dagger$  & 
\cite{Chen:2010qx,Zhao:2016ujh} & \cr
PC-F4 &  & 255.000  & 37.700  & 119.000  & 98.000  & 0.285$^\dagger$  & 0.230$^\dagger$  & 
\cite{Chen:2010qx}  & \cr
PC-LA &  & 263.000  & 37.200  & 108.000  & -61.000  & 0.268$^\dagger$  & 0.220$^\dagger$  & 
\cite{Chen:2010qx}  & \cr
PC-PK1 &  &  &  & 101.478  &  & 0.257  & 0.220  & 
\cite{Zhao:2016ujh}  & \cr
PK1 &  & 282.000  & 37.600  & 116.000  & 55.000  & 0.277  & 0.223  & 
\cite{Chen:2010qx,Zhao:2016ujh}   & \cr
PKDD &  & 263.000  & 36.900  & 90.000  & -80.000  & 0.253  & 0.214  & 
\cite{Chen:2010qx,Zhao:2016ujh}   & \cr
RAPR &  & 276.700  & 33.987  & 66.958$^\dagger$  &  & 0.201  & 0.183$^\dagger$  & 
\cite{Steiner:2004fi}   & \cr
RATP &  & 239.580  & 29.260  & 32.390  & -191.250  & 0.145$^\dagger$  & 0.152$^\dagger$  & 
\cite{Ducoin:2010as}  & \cr
rDD-ME2 &  &  &  & 51.300  &  & 0.193  & 0.179$^\dagger$  & 
\cite{RocaMaza:2011pm}   & \cr
rFSUGold &  &  &  & 60.500  &  & 0.207  & 0.186$^\dagger$  & 
\cite{RocaMaza:2011pm,Fortin:2016hny}  & \cr
rG2 &  &  &  & 100.700  &  & 0.257  & 0.214$^\dagger$  & 
\cite{RocaMaza:2011pm}  & \cr
rNL1 &  &  &  & 140.100  &  & 0.321  & 0.250$^\dagger$  & 
\cite{RocaMaza:2011pm}  & \cr
rNL3 &  &  &  & 118.500  &  & 0.280  & 0.227$^\dagger$  & 
\cite{RocaMaza:2011pm}   & \cr
rNL3* &  &  &  & 122.600  &  & 0.288  & 0.232$^\dagger$  & \cite{RocaMaza:2011pm}  & \cr
rNLC &  &  &  & 108.000  &  & 0.263  & 0.218$^\dagger$  & 
\cite{RocaMaza:2011pm}  & \cr
rNL-RA1 &  &  &  & 115.400  &  & 0.274  & 0.224$^\dagger$  & 
\cite{RocaMaza:2011pm}  & \cr
rNL-SH &  &  &  & 113.600  &  & 0.266  & 0.219$^\dagger$  & 
\cite{RocaMaza:2011pm}  & \cr
rNL-Z &  &  &  & 133.300  &  & 0.307  & 0.242$^\dagger$  & 
\cite{RocaMaza:2011pm}  & \cr
Rs &  & 237.660  & 30.593  & 80.096$^\dagger$  & -9.100  & 0.222  & 0.195$^\dagger$  & 
\cite{Steiner:2004fi,Fortin:2016hny}  & \cr
rTM1 &  &  &  & 110.800  &  & 0.271  & 0.222$^\dagger$  & 
\cite{RocaMaza:2011pm}  & \cr
R$\sigma$&  & 237.410  & 30.580  & 85.700  & -9.130  & 0.231$^\dagger$  & 0.200$^\dagger$  & 
\cite{Ducoin:2010as}  & \cr
S271 &  & 271.000  & 35.927  & 97.541$^\dagger$  &  & 0.251  & 0.211$^\dagger$  & 
\cite{Steiner:2004fi}   & \cr
SFHo &  & 245.000  & 31.600  & 47.100  &  & 0.169$^\dagger $ & 0.165$^\dagger$  & 
\cite{Oertel:2016bki}  & \cr
SFHx &  & 239.000  & 28.700  & 23.200  &  & 0.130$^\dagger$  & 0.144$^\dagger$  & 
\cite{Oertel:2016bki}  & \cr
SGI & 0.610  & 262.000  & 28.300  & 63.900  & -51.990  & 0.196$^\dagger$  & 0.180$^\dagger$  & 
\cite{Steiner:2004fi,Lim:2013tqa,Dutra:2012mb}  & \cr
SGII & 0.790  & 214.700  & 26.830  & 37.620  & -145.920  & 0.136  & 0.147$^\dagger$  & 
\cite{Ducoin:2010as,RocaMaza:2011pm,Inakura:2015cla,Dutra:2012mb} & \cr
SII & 0.580  & 341.400  & 34.160  & 50.020  & -265.720  & 0.196  & 0.177  &   
\cite{Zhao:2016ujh}  & \cr
SIII & 0.760  & 355.37 & 28.160  & 9.910  & -393.730  & 0.137  & 0.125  & 
\cite{Dutra:2012mb,Zhao:2016ujh}   & \cr
Sk$\chi$m &  & 230.400  & 30.940  & 45.600  &  & 0.167  & 0.164$^\dagger$  & 
\cite{Zhang:2017hvh,Dutra:2012mb}   & \cr
SK255 &  & 254.960  & 37.400  & 95.000  & -58.300  & 0.247$^\dagger$  & 0.208$^\dagger$  & 
\cite{Fortin:2016hny}  & \cr
SK272 &  & 271.550  & 37.400  & 91.700  & -67.800  & 0.241$^\dagger$  & 0.205$^\dagger$  & 
\cite{Fortin:2016hny}  & \cr
Ska & 0.610  & 263.160  & 32.910  & 74.620  & -78.460  & 0.214$^\dagger$  & 0.190$^\dagger$  & 
\cite{RocaMaza:2011pm,Dutra:2012mb,Fortin:2016hny,Oertel:2016bki,Inakura:2015cla,Wang:2014rva}  & \cr
Ska25-B & 0.990  & 219.000  & 32.500  & 51.000  &  & 0.176  & 0.170  & 
\cite{Brown:2013pwa}  & \cr
Ska25s20 & 0.980  & 220.750  & 33.780  & 63.810  & -118.220  & 0.196$^\dagger$  & 0.180$^\dagger$  & \cite{Dutra:2012mb}  & \cr
Ska25-T & 0.980  & 220.000  & 31.900  & 59.000  & -59.000  & 0.183  & 0.176  & 
\cite{Tsang:2019ymt}  & \cr
Ska35-B & 1.000  & 244.000  & 32.800  & 54.000  &  & 0.180  & 0.172  & 
\cite{Brown:2013pwa}  & \cr
Ska35s20 & 1.000  & 240.270  & 33.570  & 64.830  & -120.320  & 0.198$^\dagger$  & 0.181$^\dagger$  & \cite{Dutra:2012mb,Wang:2014rva}  & \cr
Ska35-T & 0.990  & 238.000  & 32.000  & 58.000  & -84.000  & 0.184  & 0.177  & 
\cite{Tsang:2019ymt,Dutra:2012mb,Ducoin:2010as}   & \cr
\hline
\end{tabular}
\end{table} 

\setcounter{table}{0}
\begin{table}[h]
\caption{Properties of 207 EoSs (3).
}
\begin{tabular}{c|ccccccc|cc}
\hline
 & m*/m & K &J & L & Ksym & Rskin-208 & Rskin-48 & Refs. \cr
\hline
SKb & 0.610  & 263.000  & 33.880  & 47.600  & -78.500  & 0.170$^\dagger$  & 0.166$^\dagger$  & 
\cite{Fortin:2016hny,Dutra:2012mb}  & \cr
SkI1 & 0.690  & 242.750  & 37.530  & 161.050  & 234.670  & 0.353$^\dagger$  & 0.268$^\dagger$  & 
\cite{Wang:2014rva,Dutra:2012mb}  & \cr
SkI2 & 0.680  & 240.700  & 33.400  & 104.300  & 70.600  & 0.262$^\dagger$  & 0.217$^\dagger$  &  \cite{Ducoin:2010as,Fortin:2016hny,Dutra:2012mb,Inakura:2015cla}   & \cr
SkI3 & 0.580  & 258.000  & 34.800  & 100.500  & 72.900  & 0.255$^\dagger$  & 0.213$^\dagger$  & 
\cite{Ducoin:2010as,Fortin:2016hny,Dutra:2012mb,Inakura:2015cla} & \cr
SkI4 & 0.650  & 247.700  & 29.500  & 60.400  & -40.600  & 0.191$^\dagger$  & 0.177$^\dagger$  & 
\cite{Ducoin:2010as,Fortin:2016hny,Dutra:2012mb,Inakura:2015cla,Lim:2013tqa}  & \cr
SkI5 & 0.580  & 255.800  & 36.697  & 129.300  & 159.500  & 0.272  & 0.214  & 
\cite{Steiner:2004fi,Ducoin:2010as,Fortin:2016hny,Dutra:2012mb,Inakura:2015cla}   & \cr
SkI6 & 0.640  & 248.650  & 30.090  & 59.700  & -47.270  & 0.189$^\dagger$  & 0.177$^\dagger$  & 
\cite{Ducoin:2010as,Fortin:2016hny,Dutra:2012mb}   & \cr
SkM* & 0.790  & 216.610  & 30.030  & 45.780  & -155.940  & 0.170  & 0.155  & 
\cite{RocaMaza:2011pm,Dutra:2012mb,Inakura:2015cla,Zhao:2016ujh}  & \cr
SkM*-B & 0.780  & 218.000  & 34.200  & 58.000  &  & 0.187  & 0.175  & 
\cite{Brown:2013pwa}  & \cr
SkM*-T & 0.790  & 219.000  & 33.700  & 65.000  & -65.000  & 0.187  & 0.179  & 
\cite{Tsang:2019ymt,} & \cr
SkMP & 0.650  & 230.930  & 29.890  & 70.310  & -49.820  & 0.197  & 0.167  & 
\cite{RocaMaza:2011pm,Ducoin:2010as,Steiner:2004f,Fortin:2016hny}  & \cr
SkO & 0.900  & 223.390  & 31.970  & 79.140  & -43.170  & 0.221$^\dagger$  & 0.194$^\dagger$  & 
\cite{Ducoin:2010as,Dutra:2012mb}   & \cr
SKOp & 0.900  & 222.360  & 31.950  & 68.940  & -78.820  & 0.204$^\dagger$  & 0.185$^\dagger$ & \cite{Dutra:2012mb,Fortin:2016hny}  & \cr
SKP & 1.000  & 200.970  & 30.000  & 19.680  & -266.600  & 0.144  & 0.144  &  
\cite{Dutra:2012mb,Zhao:2016ujh}   & \cr
SKRA & 0.750  & 216.980  & 31.320  & 53.040  & -139.280  & 0.179$^\dagger$  & 0.171$^\dagger$  & 
\cite{Dutra:2012mb}   & \cr
SKRA-B & 0.790  & 212.000  & 33.700  & 55.000  &  & 0.181  & 0.172  & 
\cite{Brown:2013pwa,Dutra:2012mb}  & \cr
SKRA-T & 0.800  & 213.000  & 33.400  & 65.000  & -55.000  & 0.190  & 0.179  & 
\cite{Tsang:2019ymt,Dutra:2012mb}   & \cr
Sk-Rs &  &  &  & 85.700  &  & 0.215  & 0.191$^\dagger$  & 
\cite{RocaMaza:2011pm}  & \cr
SkSM* &  &  &  & 65.500  &  & 0.197  & 0.181$^\dagger$  & 
\cite{RocaMaza:2011pm}  & \cr
SkT1 & 1.000  & 236.160  & 32.020  & 56.180  & -134.830  & 0.184$^\dagger$  & 0.173$^\dagger$  &   
\cite{Dutra:2012mb}   & \cr
SkT1-B & 0.970  & 242.000  & 33.300  & 56.000  &  & 0.183  & 0.172  & 
\cite{Brown:2013pwa}  & \cr
SkT1-T & 0.970  & 238.000  & 32.600  & 63.000  & -70.000  & 0.190  & 0.179  & 
\cite{Tsang:2019ymt,Dutra:2012mb}   & \cr
SkT2 & 1.000  & 235.730  & 32.000  & 56.160  & -134.670  & 0.184$^\dagger$  & 0.173$^\dagger$  &   
\cite{Dutra:2012mb}   & \cr
SkT2-B & 0.970  & 242.000  & 33.500  & 58.000  &  & 0.186  & 0.174  & 
\cite{Brown:2013pwa}  & \cr
SkT2-T & 0.960  & 238.000  & 32.600  & 62.000  & -75.000  & 0.188  & 0.178  & 
\cite{Tsang:2019ymt,Dutra:2012mb}   & \cr
SkT3 & 1.000  & 235.740  & 31.500  & 55.310  & -132.050  & 0.182$^\dagger$  & 0.173$^\dagger$  &  
\cite{Dutra:2012mb}   & \cr
SkT3-B & 0.980  & 241.000  & 32.700  & 53.000  &  & 0.179  & 0.172  & 
\cite{Brown:2013pwa}  & \cr
SkT3-T & 0.970  & 236.000  & 31.900  & 58.000  & -80.000  & 0.183  & 0.178  & 
\cite{Tsang:2019ymt}  & \cr
Sk-T4 & 1.000  & 235.560  & 35.457  & 94.100  & -24.500  & 0.253  & 0.212$^\dagger $ & 
\cite{Steiner:2004fi,Inakura:2015cla,RocaMaza:2011pm}  & \cr
Sk-T6 & 1.000  & 235.950  & 29.970  & 30.900  & -211.530  & 0.151  & 0.155$^\dagger $ & 
\cite{RocaMaza:2011pm,Dutra:2012mb}   & \cr
Skxs20 & 0.960  & 201.950  & 35.500  & 67.060  & -122.310  & 0.201$^\dagger$  & 0.183$^\dagger$  & \cite{Dutra:2012mb}   & \cr
Skz2 & 0.700  & 230.070  & 32.010  & 16.810  & -259.660  & 0.120$^\dagger$  & 0.138$^\dagger$  &  \cite{Dutra:2012mb,Wang:2014rva}   & \cr
Skz4 & 0.700  & 230.080  & 32.010  & 5.750  & -240.860  & 0.102$^\dagger$  & 0.128$^\dagger$  &   \cite{Dutra:2012mb,Wang:2014rva}   & \cr
SLy0 & 0.700  & 229.670  & 31.982  & 44.873$^\dagger$  & -116.230  & 0.165  & 0.163$^\dagger$  & 
\cite{Steiner:2004fi,Dutra:2012mb}   & \cr
SLy10 & 0.680  & 229.740  & 31.980  & 38.740  & -142.190  & 0.155$^\dagger$  & 0.158$^\dagger$  & 
\cite{Ducoin:2010as,Dutra:2012mb}   & \cr
Sly2 & 0.700  & 229.920  & 32.000  & 47.460  & -115.130  & 0.170$^\dagger$  & 0.166$^\dagger$ & \cite{Dutra:2012mb,Fortin:2016hny}   & \cr
SLy230a & 0.700  & 229.890  & 31.980  & 44.310  & -98.210  & 0.155  & 0.158$^\dagger$  & 
\cite{Steiner:2004fi,Ducoin:2010as,Fortin:2016hny,Dutra:2012mb}    & \cr
Sly230b & 0.690  & 229.960  & 32.010  & 45.960  & -119.720  & 0.167$^\dagger$  & 0.164$^\dagger$  & \cite{Ducoin:2010as,Dutra:2012mb}   & \cr
SLy4 & 0.690  & 229.900  & 32.000  & 45.900  & -119.700  & 0.162  & 0.152  & 
\cite{RocaMaza:2011pm,Lim:2013tqa,Inakura:2015cla,Gonzalez-Boquera:2017rzy,Fortin:2016hny,Zhao:2016ujh}   & \cr
SLy4-B & 0.700  & 224.000  & 34.100  & 56.000  &  & 0.184  & 0.174  & 
\cite{Brown:2013pwa}  & \cr
SLy4-T & 0.760  & 222.000  & 33.600  & 66.000  & -55.000  & 0.191  & 0.179  & 
\cite{Tsang:2019ymt,Dutra:2012mb}   & \cr
SLy5 & 0.700  & 229.920  & 32.010  & 48.150  & -112.760  & 0.162  & 0.160  &
\cite{Dutra:2012mb,Zhao:2016ujh}   & \cr
SLy6 & 0.690  & 229.860  & 31.960  & 47.450  & -112.710  & 0.161  & 0.152   &
\cite{Dutra:2012mb,Zhao:2016ujh}   & \cr
Sly9 & 0.670  & 229.840  & 31.980  & 54.860  & -81.420  & 0.182$^\dagger$  & 0.172$^\dagger$  & \cite{Dutra:2012mb,Fortin:2016hny}   & \cr
SQMC650 & 0.780  & 218.110  & 33.650  & 52.920  & -173.150  & 0.178$^\dagger$  & 0.171$^\dagger$  & \cite{Dutra:2012mb}   & \cr
SQMC700 & 0.760  & 222.200  & 33.470  & 59.060  & -140.840  & 0.188$^\dagger$  & 0.176$^\dagger$  & \cite{Dutra:2012mb}   & \cr
SQMC750-B & 0.710  & 228.000  & 34.800  & 59.000  &  & 0.190  & 0.176  & 
\cite{Brown:2013pwa}  & \cr
SQMC750-T & 0.750  & 223.000  & 33.900  & 68.000  & -50.000  & 0.194  & 0.180  & 
\cite{Tsang:2019ymt,Dutra:2012mb}   & \cr
SR1 & 0.900  & 202.150  & 29.000  & 41.245$^\dagger$  &  & 0.160  & 0.160$^\dagger$  & 
\cite{Steiner:2004fi}  & \cr
SR2 &  & 224.640  & 30.071  & 49.130$^\dagger$  &  & 0.172  & 0.167$^\dagger$  & 
\cite{Steiner:2004fi}   & \cr
SR3 &  & 222.550  & 29.001  & 48.308$^\dagger$  &  & 0.171  & 0.166$^\dagger$  & 
\cite{Steiner:2004fi}   & \cr
SV & 0.380  & 306.000  & 32.800  & 96.100  & 24.190  & 0.230  & 0.196  & 
\cite{Lim:2013tqa,Ducoin:2010as,Zhao:2016ujh,Dutra:2012mb}    & \cr
SV-bas & 0.900  & 221.760  & 30.000  & 32.000  & -156.570  & 0.155  & 0.158$^\dagger$  & 
\cite{Reinhard:2016sce,Dutra:2012mb}  & \cr
SV-K218 & 0.900  & 218.230  & 30.000  & 35.000  & -207.870  & 0.161  & 0.161$^\dagger$  & 
\cite{Reinhard:2016sce,Dutra:2012mb}  & \cr
SV-K226 & 0.900  & 225.820  & 30.000  & 34.000  & -211.920  & 0.159  & 0.160$^\dagger$  & 
\cite{Reinhard:2016sce,Dutra:2012mb}   & \cr
SV-K241 & 0.900  & 241.070  & 30.000  & 31.000  & -230.770  & 0.151  & 0.155$^\dagger$  & 
\cite{Reinhard:2016sce,Dutra:2012mb}  & \cr
SV-kap00 & 0.900  & 233.440  & 30.000  & 40.000  & -161.780  & 0.158  & 0.159$^\dagger$  & 
\cite{Reinhard:2016sce,Dutra:2012mb}  & \cr
SV-kap20 & 0.900  & 233.440  & 30.000  & 36.000  & -193.190  & 0.155  & 0.158$^\dagger$  & 
\cite{Reinhard:2016sce,Dutra:2012mb}  & \cr
SV-kap60 & 0.900  & 233.450  & 30.000  & 29.000  & -249.750  & 0.154  & 0.157$^\dagger$  & 
\cite{Reinhard:2016sce,Dutra:2012mb}   & \cr
SV-L25 & 0.900  &  & 30.000  & 25.000  &  & 0.143  & 0.151$^\dagger$  & 
\cite{Reinhard:2016sce}   & \cr
SV-L32 & 0.900  &  & 30.000  & 32.000  &  & 0.154  & 0.157$^\dagger$  & 
\cite{Reinhard:2016sce}  & \cr
SV-L40 & 0.900  & 233.3 & 30.000  & 40.000  &  & 0.166  & 0.164$^\dagger$  & 
\cite{Reinhard:2016sce}  & \cr
SV-L47 & 0.900  & 233.4 & 30.000  & 47.000  &  & 0.177  & 0.170$^\dagger$  & 
\cite{Reinhard:2016sce}  & \cr
\hline
\end{tabular}
\end{table}

\setcounter{table}{0}
\begin{table}[h]
\caption{Properties of 207 EoSs (4).
}
\begin{tabular}{c|ccccccc|cc}
\hline
 & m*/m & K &J & L & Ksym & Rskin-208 & Rskin-48 & Refs. \cr
\hline
SV-mas07 & 0.700  & 233.540  & 30.000  & 52.000  & -98.770  & 0.152  & 0.156$^\dagger$  & 
\cite{Reinhard:2016sce,Dutra:2012mb}  & \cr
SV-mas08 & 0.800  & 233.130  & 30.000  & 40.000  & -172.380  & 0.160  & 0.160$^\dagger$  & 
\cite{Reinhard:2016sce,Dutra:2012mb,Wang:2014rva}  & \cr
SV-mas10 & 1.000  & 234.330  & 30.000  & 28.000  & -252.500  & 0.152  & 0.156$^\dagger$  & 
\cite{Reinhard:2016sce,Dutra:2012mb}  & \cr
SV-sym28 & 0.900  & 240.860  & 28.000  & 7.000  & -305.940  & 0.117  & 0.136$^\dagger$  & 
\cite{Reinhard:2016sce,Dutra:2012mb}  & \cr
SV-sym32 & 0.900  & 233.810  & 32.000  & 57.000  & -148.790  & 0.192  & 0.178$^\dagger$  & 
\cite{Reinhard:2016sce,Dutra:2012mb}  & \cr
SV-sym32-B & 0.910  & 237.000  & 32.300  & 51.000  &  & 0.176  & 0.174  & 
\cite{Brown:2013pwa}  & \cr
SV-sym32-T & 0.910  & 232.000  & 31.500  & 58.000  & -77.000  & 0.181  & 0.179  & 
\cite{Tsang:2019ymt}  & \cr
SV-sym34 & 0.900  & 234.070  & 34.000  & 81.000  & -79.080  & 0.227  & 0.198$^\dagger$  &   
\cite{Reinhard:2016sce,Dutra:2012mb,Wang:2014rva}  & \cr
TFa &  & 245.100  & 35.050  & 82.500  & -68.400  & 0.250  & 0.210$^\dagger$  & 
\cite{Fattoyev:2013yaa} & \cr
TFb &  & 250.100  & 40.070  & 122.500  & 45.800  & 0.300  & 0.238$^\dagger$  & 
\cite{Fattoyev:2013yaa}  & \cr
TFc &  & 260.500  & 43.670  & 135.200  & 51.600  & 0.330  & 0.255$^\dagger$  & 
\cite{Fattoyev:2013yaa} & \cr
TM1 &  & 281.000  & 36.900  & 110.800  & 33.550  & 0.272$^\dagger$  & 0.223$^\dagger$  & 
\cite{Ishizuka:2014jsa,Ducoin:2010as,Fortin:2016hny,Chen:2010qx}  & \cr
TW99 &  & 241.000  & 32.800  & 55.000  & -124.000  & 0.196  & 0.186  &  
\cite{Chen:2010qx,Zhao:2016ujh}  &\cr 
UNEDF0 &  & 229.800  & 30.500  & 45.100  & -189.600  & 0.166$^\dagger$  & 0.164$^\dagger$  & 
\cite{Inakura:2015cla,Dutra:2012mb} &\cr 
UNEDF1 &  & 219.800  & 29.000  & 40.000  & -179.400  & 0.158$^\dagger$  & 0.159$^\dagger$  & 
\cite{Inakura:2015cla}  & \cr
Z271 &  & 271.000  & 35.369  & 89.520$^\dagger$  &  & 0.238  & 0.204$^\dagger$  & 
\cite{Steiner:2004fi}   & \cr
Sly7 &   &       &     &   45.900 &  & 0.150  & 0.153$$  & 
\cite{Matsuzaki:2022spe,Matsuzaki:2021hdm}   & \cr

\hline
\end{tabular}
\end{table}   


\end{document}